\documentclass[onecolumn,aps,prl]{revtex4-2}
\usepackage{epsf}
\usepackage{graphicx}
\usepackage{sidecap}
\usepackage{amsmath}
\usepackage{float}
\usepackage{hyperref}
\hypersetup{
	colorlinks=true,
	citecolor=red,
	linkcolor=blue,
	filecolor=blue,   
	urlcolor=blue,
}
\renewcommand{\thefigure}{{\textbf{\arabic{figure}}}}

\begin{document}
	
	\title{\Large{Observation of paramagnetic spin-degeneracy lifting in EuZn$_2$Sb$_2$}}%
	
	\author{\large{Milo~X.~Sprague$^1$, Sabin~Regmi$^{1,2}$, Barun~Ghosh$^{2,6}$, Anup~Pradhan~Sakhya$^1$, Mazharul~Islam~Mondal$^1$, Iftakhar~Bin~Elius$^1$, Nathan~Valadez$^1$, Bahadur Singh$^3$, Tetiana Romanova$^4$, Dariusz~Kaczorowski$^4$, Arun~Bansil$^{2,6}$, Madhab~Neupane$^{1}$}} \thanks{Corresponding author:\href{mailto:madhab.neupane@ucf.edu} {madhab.neupane@ucf.edu}}
	
	\affiliation{\textcolor{white}{Secret Text!}\\$^1$Department of Physics, University of Central Florida, Orlando, Florida 32816, USA\\$^2$Center for Quantum Actinide Science and Technology, Idaho National Laboratory, Idaho Falls, Idaho 83415, USA\\$^3$Department of Physics, Northeastern University, Boston, Massachusetts 02115, USA\\$^4$Department of Condensed Matter Physics and Materials Science, Tata Institute of Fundamental Research, Mumbai 400005, India\\$^5$Institute of Low Temperature and Structure Research, Polish Academy of Sciences, ul. Okólna 2, PL-50-422 Wroc\l aw, Poland\\$^6$Quantum Materials and Sensing Institute, Northeastern University, Burlington, Massachusetts 01803, USA\\}

\maketitle
\clearpage

\section{abstract}
	Taken together, time-reversal and spatial inversion symmetries impose a two-fold spin degeneracy of the electronic states in crystals. In centrosymmetric materials, this degeneracy can be lifted by introducing magnetism, either via an externally applied field or through internal magnetization. However, a correlated alignment of spins, even in the paramagnetic phase, can lift the spin degeneracy of electronic states. Here, we report an in-depth study of the electronic band structure of the Eu-ternary pnictide EuZn$_2$Sb$_2$ through a combination of high-resolution angle-resolved photoemission spectroscopy measurements and first principles calculations. An analysis of the photoemission lineshapes over a range of incident photon energies and sample temperatures is shown to reveal the presence of band spin degeneracy-lifting in the paramagnetic phase. Our ARPES results are in good agreement with theoretical ferromagnetic-phase calculations, which indicates the importance of ferromagnetic fluctuations in the system. Through our calculations, we predict that spin-polarized bands in EuZn$_2$Sb$_2$ generate a single pair of Weyl nodes. Our observation of band-splitting in EuZn$_2$Sb$_2$ provides a key step toward realizing time-reversal symmetry breaking physics in the absence of long-range magnetic order.

	\section{I. Introduction} 
	Symmetry has become a significant organizing principle in modern condensed matter physics. Symmetries enforce protected degeneracies along high-symmetry directions in the Brillouin zone (BZ), which can dramatically simplify the analysis of electronic band structures \cite{grouptheory}. Consideration of symmetries have also led to efficient classification schemes for topologically protected states \cite{symmetryclasstop}. Topological insulators provided the first example of a three-dimensional (3D) symmetry protected topological material \cite{fu2007topological, hsieh2008topological, xia2009observation, zhang2009topological}. Time-reversal symmetry (TRS) and bulk-inversion symmetry (IS) were shown to protect spin-orbit induced band inversions and two-fold band crossings between spin-polarized Dirac surface states pinned to the TRS points of the BZ \cite{fu2007topological, hasan2010colloquium, qi2011topological, bansil2016topological}. These symmetries can also protect the 4-fold degenerate band crossings in bulk-3D Dirac semimetals \cite{young2012dirac,wang2012dirac,neupane2014observation, armitage2018weyl}. The breaking of TRS in topological insulators and Dirac semimetals can lead to new topological phases such as the magnetic topological insulators \cite{tokura2019magnetic} and the Weyl semimetals \cite{armitage2018weyl, yan2017topological}.\\
		
	Effects of TRS breaking are usually explored in the context of materials with net magnetization. However, short-ranged correlated spin alignments can provide another mechanism for lifting the Kramers degeneracy \cite{sokoloff1975fluctuation}. Given large enough correlation lengths and times, it can be anticipated that similar TRS-breaking physics may be observed in materials \cite{ma2019spin}. Zintl-phase Eu$X_2Pn_2$, with transition metal $X$ and pnictogen $Pn$, most notably EuCd$_2$As$_2$ and EuZn$_2$As$_2$, have been recognized for supporting nonlinear anomalous Hall and anomalous Nernst effects \cite{xu2021unconventional,yi2023topological}, electron spin resonance (ESR) \cite{ma2019spin, goryunov2012esr,goryunov2014spin}, muon spin relaxation \cite{ma2019spin}, and resonant elastic X-ray scattering \cite{soh2020resonant} responses that are consistent with the formation of slowly fluctuating ferromagnetic domains in the paramagnetic phase. These results indicate the possibility of producing spin-polarized electronic states in paramagnetic samples, where both the global TRS and IS are preserved. The low-energy physics of Eu$X_2Pn_2$ systems are dominated by the interplay of transition-metal-$s$ conduction bands and pnictide-$p$ valence bands near the $\Gamma$-point \cite{wang2019single}. The precise energetics are strongly dependent on spin-orbit coupling strength and details of the resulting magnetic ordering. This interplay of magnetism and spin-orbit coupling stabilizes a variety of magnetic and non-magnetic topological phases \cite{xu2019higher, sato2020signature, regmi2020temperature, ma2019spin, wang2019single, KaczorowskiEuZn2Sb2, wang2023intrinsic, SpragueEuSnP, hua2018dirac,gui2019new, li2019dirac, jo2020manipulating}.
	\\
	
	The recent interest in EuZn$_2$As$_2$ and EuCd$_2$As$_2$ naturally motivates continuing work on EuZn$_2$Sb$_2$ due to its intermediate spin-orbit coupling strength. EuZn$_2$Sb$_2$ has been observed to produce ESR lineshapes consistent with slow ferromagnetic correlations ranging from the antiferromagnetic (AFM) transition temperature ($T_N$=13 K) to over 100 K \cite{goryunov2012esr}.  Given the connection between magnetic fluctuations and electronic structure established both theoretically \cite{sokoloff1975fluctuation} and through recent experiments \cite{xu2021unconventional, yi2023topological, ma2019spin, goryunov2012esr, goryunov2014spin, soh2020resonant}, the potential interplay between magnetic correlations and spin-orbit coupling induced topology in EuZn$_2$Sb$_2$ warrants a detailed investigation into the electronic structure of this material. We have carried out angle-resolved photoemission spectroscopy (ARPES) measurements along with parallel density functional theory (DFT) based calculations. We report direct observation of band spin-degeneracy lifting through an analysis of ARPES lineshapes over a range of incident photon energies and sample temperatures in the paramagnetic phase. We observe asymmetric ARPES lineshapes that are consistent with the convolution of two distinct band peaks, which persist over a wide temperature range of 15 K and 130 K. Our measurements show good agreement with our DFT predictions for the ferromagnetic phase below the Fermi level. Interestingly, our calculations for the spin-split bands are predicted to extend above the Fermi level to generate a single pair of Weyl nodes separated along the k$_z$-direction. \\

	\begin{figure*}
		\centering
		\includegraphics[width=0.9\linewidth]{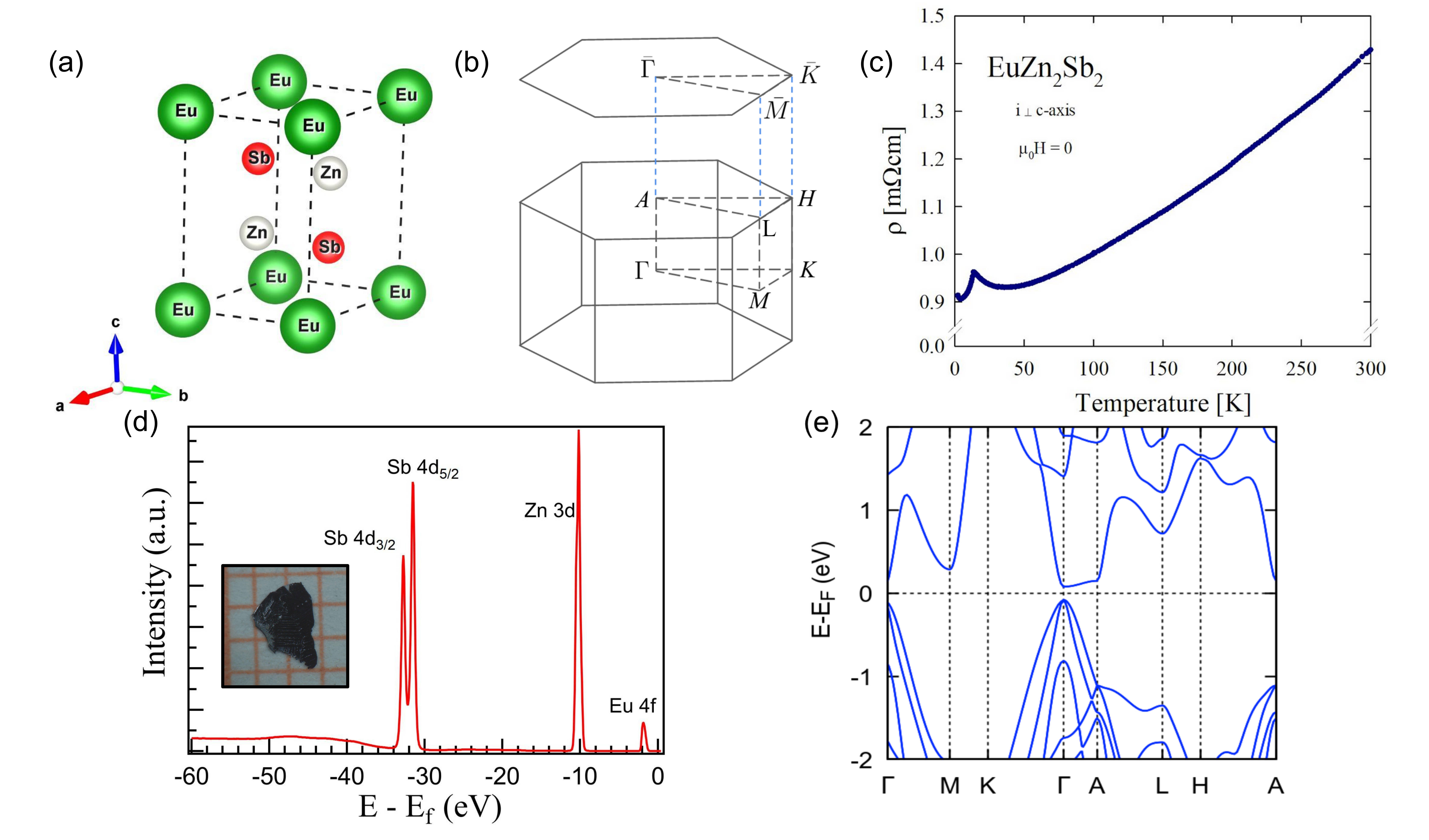}
		\caption{Crystal structure and sample characterization of EuZn$_2$Sb$_2$. (a) Crystal structure of EuZn$_2$Sb$_2$. The green, gray, and red balls identify Eu, Zn, and Sb atoms, respectively. (b) The associated bulk BZ and the surface BZ projected along (001) with high symmetry points marked. (c) Temperature dependence of the electrical resistivity in zero applied magnetic field measured within the crystallographic ab-plane. (d) Core-level photoemission spectrum with characteristic peaks of Eu 4$f$, Zn 3$d$, and Sb 4$d$ orbitals. A sample photograph is shown in the inset. (e) Non-magnetic band dispersion calculation along several high-symmetry directions (with taking into account SOC).}
		\label{fig:eu12sb2-figure-1}
	\end{figure*}

	\section{II. Methods}
	High-quality EuZn$_2$Sb$_2$ single crystals were grown by using Zn-Sb flux and characterized using X-ray diffraction and energy-dispersive X-ray spectroscopy (see the Supplemental Material (SM) for growth and characterization details). High-resolution ARPES measurements were carried out at the Stanford Synchrotron Radiation Laboratory (SSRL) beamline 5-2 and at the Advanced Light Source (ALS) beamline 4.0.3, equipped with Scienta DA30 analyzer and Scienta R8000 electron analyzer, respectively (see SM for further details). All electronic structure calculations were performed within the DFT framework using a plane wave basis set in the Vienna Ab initio Simulation Package (VASP) \cite{kresse1996efficient}. The band structures shown in the main text are calculated using the Heyd-Scuseria-Ernzerhof (HSE) functional \cite{kresse1999ultrasoft} (see SM for computational details).\\ 
	
	\section{III. Results}
	
	\subsection{III.i. Crystal Structure and Electronic Properties}
	EuZn$_2$Sb$_2$ crystallizes in the trigonal space group $P\overline{3}m1$ (No. 164). The unit cell structure consists of alternating Zn$_2$Sb$_2$ and Eu layers, as depicted in FIG.~1(a) \cite{weber2006low}. A hexagonal bulk BZ is produced by the trigonal lattice (shown in FIG.~1(b)). This crystal structure has a natural cleave along the (001) plane, which produces the surface-projected BZ shown in relation to the bulk. Previous work on Eu$X_2Pn_2$ samples shows a strong sensitivity of the carrier concentration and Fermi level position on details of crystal synthesis protocol \cite{santos2023eucd}. Resistivity measurements were performed to check whether the low-energy physics is described by metallic, semimetallic, or semiconducting behavior. FIG. 1(c) displays the temperature dependence of the electrical resistivity in single-crystalline EuZn$_2$Sb$_2$. The sample exhibits metallic behavior with a pronounced anomaly at the antiferromagnetic phase transition in the form of a cusp-like peak in the resistivity. The N\'eel temperature, T$_N$ = 13.3 K, is close to the value previously reported \cite{weber2006low,zhang2008new}. Notably, the magnitude of $\rho$(T) is rather large, and the resistivity variation on decreasing temperature is fairly small, which indicates that EuZn$_2$Sb$_2$ is a bad metal. A negative temperature coefficient above $T_N$ likely arises due to short range magnetic interactions, while a decrease of the resistivity below $T_N$ is clearly a result of the reduction in electron scattering on the magnetic moments in the ordered state. The short-range magnetic interactions observed above $T_N$ bear a ferromagnetic character, as indicated by the positive value of the paramagnetic Curie temperature derived from the magnetic susceptibility data (see SM). The elemental composition of the sample is corroborated by the core-level photoemission results, which show the presence of atomic peaks at Eu 4$f$, Zn 3$d$, and Sb 4$d$ binding energies (FIG.~1(d)). \\

	\begin{figure*}
		\centering
		\includegraphics[width=1\linewidth]{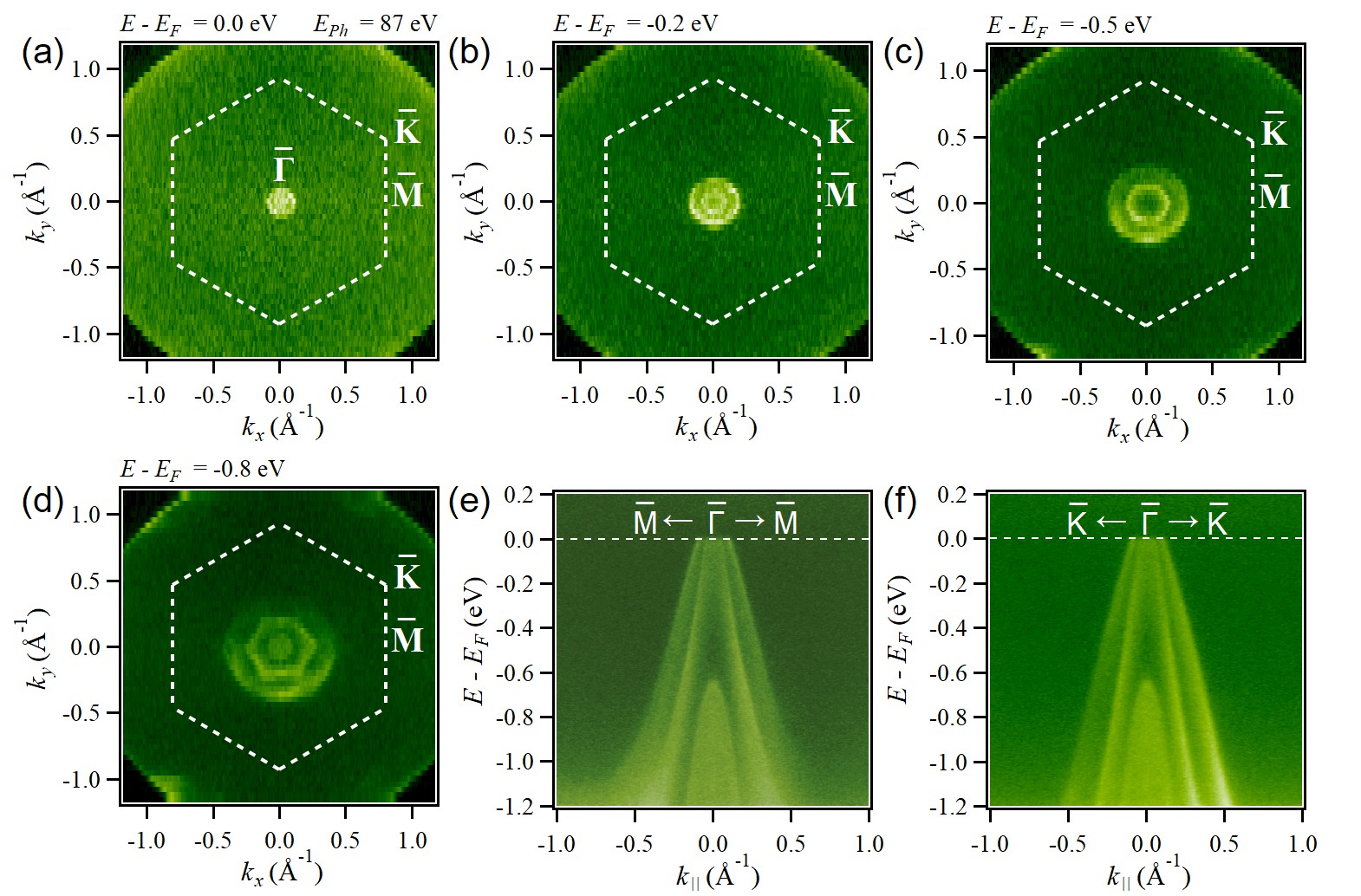}
		\caption{Constant energy contours (CECs) and dispersion cuts taken with an incident photon energy of 86 eV. (a) The Fermi surface and two CECs at binding energies of (b) 200 meV, (c) 500 meV, and (d) 800 meV. Dispersion cuts along the (e) $\overline{\mathrm{M}}-\overline{\Gamma}-\overline{\mathrm{M}}$ and (f) $\overline{\text{K}}-\overline{\Gamma}-\overline{\mathrm{K}}$  high-symmetry directions.}
		\label{fig:eu12sb2-figure-2-12-29-2022}
	\end{figure*}

	In disagreement with the metallic behavior found in electrical resistivity measurements, our non-magnetic calculations (FIG. 1(e)) predict a direct semiconducting band gap at the $\Gamma$-point (FIG 1.(e)). To better understand the valence electronic structure of EuZn$_2$Sb$_2$, we have performed high-resolution ARPES, the results of which are shown in FIGs.~2 and 3. The Fermi surface and several constant energy contours (CECs) are shown in FIG.~2(a-d). The Fermi surface (FIG.~2(a)) consists of two circular hole pockets of small area localized at the $\overline{\Gamma}$-point. The hole nature of the associated bands is evident from the increase in size of the pocket with increasing binding energy. The dispersion cuts taken along the $\mathrm{\overline{M}}-\mathrm{\overline{\Gamma}}-\mathrm{\overline{M}}$ (FIG.~2(e)) and $\mathrm{\overline{K}}-\mathrm{\overline{\Gamma}}-\mathrm{\overline{K}}$ (FIG.~2(f)) directions show that the two hole bands disperse in parallel, where they are joined by the maximum of a third parabolic band at a higher binding energy. The two cuts show a similar dispersion behavior close to the Fermi level, only deviating from one another at higher binding energies.\\

	\begin{figure*}
		\centering
		\includegraphics[width=1\linewidth]{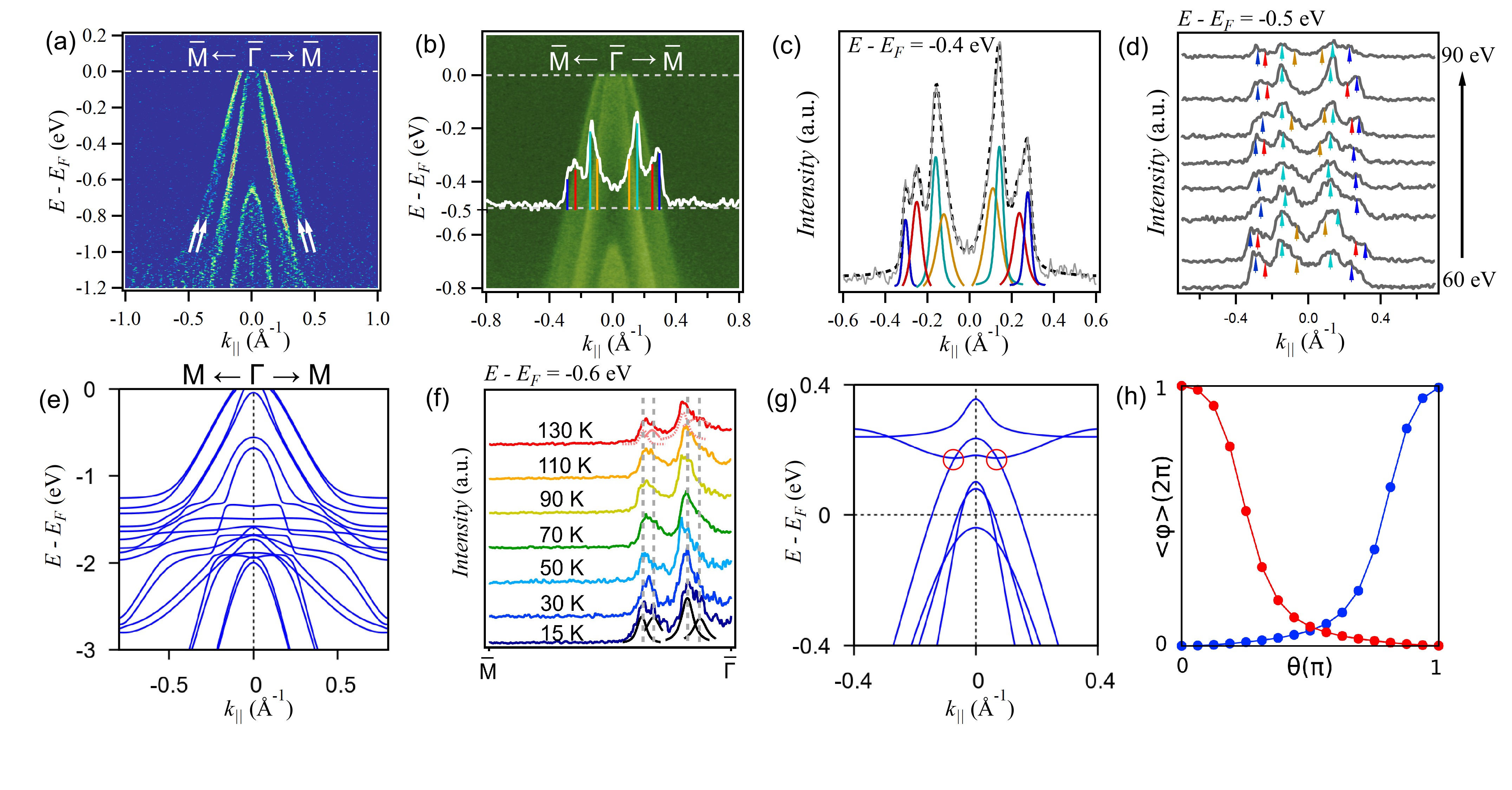}
		\caption{Observation of spin-split bands in EuZn$_2$Sb$_2$. SD edge detections applied to the MDCs and displayed for the (a) $\mathrm{\overline{M}}-\mathrm{\overline{\Gamma}}-\mathrm{\overline{M}}$ direction. (b) MDC at 0.5 eV binding energy integrated over an 8 meV window displayed over the $\overline{\mathrm{M}}-\overline{\Gamma}-\overline{\mathrm{M}}$ dispersion cut. Vertical black lines indicate fitted peak positions. (c) Results of Voigt function fit to the MDC at 0.4 eV binding energy. The total fit is given in blue, the individual peaks in green, and the raw MDC in red. (d) MDCs stacked over incident photon energies. Taken along the $\mathrm{\overline{M}}-\mathrm{\overline{\Gamma}}-\mathrm{\overline{M}}$ direction at a binding energy of 0.5 eV. 8 stacked MDCs for photon energies ranging between 62 and 90 eV corresponding to those in FIG. 2S. Vertical arrows indicate the peak position resulting from the peak fitting for each MDC. The colors indicating the position of the fitted peaks are made consistent between panels (b-d), where blue/red and teal/orange indicates the spin-split pair of the outermost bands and innermost bands, respectively. (e) Result of our ferromagnetic DFT calculations for an out-of-plane magnetization along  $\mathrm{M}-\mathrm{\Gamma}-\mathrm{M}$. (f) Temperature dependence of the 0.6 eV binding energy MDC along $\mathrm{\overline{M}}-\mathrm{\overline{\Gamma}}-\mathrm{\overline{M}}$ for temperatures between 15 K and 130 K. Fitted peaks are shown at 15 K and 130 K. (g) Out-of-plane ferromagnetic  calculations above the Fermi level showing the presence of Weyl nodes 180 meV above the Fermi level (circled in red). (h) The chirality of the Weyl points, calculated using the Wannier charge center evolution around a sphere enclosing the gapless points. The red and blue curves are for the two different Weyl points. The Weyl points have chirality $\pm{1}$.}
		\label{fig:eu12sb2-figure-3-01-31-2023}
	\end{figure*}

\subsection{III.ii. Band Splitting Induced Weyl Semimetallic Phase}
	A close inspection of the dispersion cuts reveal the presence of low intensity parallel dispersing bands, which are not present in the non-magnetic DFT calculations, as can be seen in FIG 1(e). To better visualize these split bands, second derivative (SD) edge detection method was applied to the $\mathrm{\overline{M}}-\mathrm{\overline{\Gamma}}-\mathrm{\overline{M}}$ dispersion cut  (FIG.~3(a)). The SD figures resolve these features as distinct peaks in the photoemission intensity. In the raw photoemission data, however, these peaks constructively add together to produce an asymmetric momentum distribution curve (MDC) lineshape (FIG.~3(b-d)). We performed a curve fit on the MDC using a Voigt function, which is the convolution of the idealized Lorentzian lineshape with Gaussian broadening to account for resolution and temperature effects \cite{levy2014deconstruction}. Fitted results are shown in FIGs. 3(b-d), where each peak is color-coded such that the blue/red and teal/orange corresponds to the spin-split pairs from the outer and inner bands, respectively. Agreement with the measured MDCs was obtained by including two closely dispersing band pairs, giving an indication of band splitting. The peak positions are indicated for the MDC taken along the $\mathrm{\overline{M}}-\mathrm{\overline{\Gamma}}-\mathrm{\overline{M}}$ cut at a binding energy of 0.4 eV. The seperation of the outer two peaks is about 0.05 $\pm$ 0.004 $\text{\AA}^{-1}$ and about 0.03 $\pm$ 0.004 $\text{\AA}^{-1}$ for the inner bands, as obtained from the difference in the fitted peak positions. The ARPES spectra shown in FIG. 3(a-c) were taken at a photon energy of $h\nu =$ 86 eV. To demonstrate the persistence of the band-splitting we across various photon energies, we have plotted the MDCs for the same $\mathrm{\overline{M}}-\mathrm{\overline{\Gamma}}-\mathrm{\overline{M}}$ cut for various photon energies ranging from 60 eV to 90 eV at $E-E_F$ = -0.5 eV binding energy, as shown in FIG. 3(d).\\
	\indent Non-magnetic calculations assign a spin-degeneracy to all the bands, including the two outer-most hole pockets where spin-splitting is observed experimentally (FIG 3). In accounting for the potential split-bands, we repeated the DFT calculation in an out-of-plane ferromagnetic configuration, which produce the pair of non-degenerate parallel dispersive bands (FIG.~3(e)). The projected size of the band-splitting is rather small for the highly dispersive band energies, although the comparison with experimental results establish the presence of ferromagnetic interactions that result in the splitting of bands. Similar observations in related materials have been attributed to short-range ferromagnetic fluctuations in the paramagnetic phase \cite{ma2019spin}. Given the presence of slow ferromagnetic correlations ranging from the N\'eel temperature to over 100 K in EuZn$_2$Sb$_2$ \cite{goryunov2012esr}, a similar mechanism can be expected for the observation of the band-splitting feature in our measurements. We repeated measurements of the $\mathrm{\overline{M}}-\mathrm{\overline{\Gamma}}-\mathrm{\overline{M}}$ dispersion cut over sample temperatures between 15~K and 130~K. The temperature variations of the MDCs taken at a binding energy of 0.5 eV are stacked in FIG.~3(f). Weak temperature dependence is seen in the MDC lineshapes, which indicates the presence of band splitting up to 130 K, which is consistent with previous reports of the presence of weak ferromagnetic correlations over 100 K \cite{goryunov2012esr}. These results show that, even if EuZn$_2$Sb$_2$ is paramagnetic above 13 K, the presence of weak ferromagnetic domains lifts the spin degeneracy of bands and leads to the observed splitting.\\

	While our non-magnetic calculations predict a semiconducting gap, our results show two semimetallic hole pockets comprising the Fermi surface. Decreasing the Fermi level by about 180 meV in both the non-magnetic and ferromagnetic calculations replicates our ARPES results, even though only the ferromagnetic calculations correctly account for the band-splitting seen experimentally. Interestingly, our ferromagnetic calculations predict the vanishing of this gap in the spin-polarized state, which extend to produce a single pair of Weyl nodes along the $\mathrm{A}-\mathrm{\Gamma}-\mathrm{A}$ direction (FIG.~3(g)). After accounting for the Fermi level shift in our calculations, possibly due to hole-doping in our samples, the Weyl points are predicted to lie around 180 meV above the Fermi level. We confirm the nontrivial chirality of the Weyl crossings by performing Wannier charge center evolution calculations around a sphere enclosing the gapless point, the results of which are shown in FIG.~3(h). The simultaneous stabilization of the ferromagnetic ground state and electron-doping would make EuZn$_2$Sb$_2$ a candidate minimal Weyl semimetal.\\
	
	The observation of band-splitting in a centrosymmetric and non-magnetic system may stem from various sources alternate to what has been proposed here. The first possibility is that the observed splitting is surface originated. The spin-degeneracy of generic momentum points is in general lifted at the surface due to 3D inversion symmetry breaking \cite{Manchon2015Rashba}, which may allow spin-orbit interactions to lift the spin-degeneracy. A second explanation for the band splitting is through the influence of crystal defects and domain boundaries. Surface reconstructions could induce modifications in the electronic states near the surface, contributing to the observed band-splitting phenomenon. The influence of surface reconstructions are highly detectable through ARPES measurements through a reduction of the reciprocal space periodicity \cite{Liu2021Surface}. However, this scenario contrasts with the data presented in FIG. 2, where no additional periodicity of the bands is observed within the first BZ. Lastly, it is essential to consider the possibility of artifacts caused by $k_z$-broadening, which may arise due to the limited photon escape depth and the presence of multiple photoelectron scattering events \cite{Strocov2003Kperp}. The issue of $k_z$-broadening is an ever-present limitation in VUV ARPES experiments. Such broadening generally leads to an asymmetric lineshape in the momentum distribution curves, becoming particularly pronounced at the extreme points of the $k_z$ dispersion. Despite this, we attribute the observed ARPES spectra to the presence of band-splitting. The influence of $k_z$ broadening is strongly photon-dependent, as modifying the incident photon energy changes the central $k_z$ plane being measured. However, the apparent band-splitting does not exhibit substantial variation with incident photon energy, despite different regions of the valence band being emphasized along $k_z$.

	\section{IV. Conclusions}
	Eu-ternary pnictides have been recognized for their simple fermiology, magnetic ordering, as well for their ability to host a non-trivial band topology. Motivated by the recent evidence of local ferromagnetic behavior in the macroscopic paramagnetic phase of EuZn$_2$Sb$_2$, we report high-resolution ARPES and parallel ab initio calculations to unravel how magnetic fluctuations affect the electronic band structure of this material. We adduce evidence that the paramagnetic band spin-degeneracy is lifted via an analysis of the MDC lineshapes, which is found to display an asymmetric deviation from the expected Voigt function based lineshape. This asymmetric lineshape is consistent with the spin-splitting of electronic states involving slow ferromagnetic fluctuations in the paramagnetic phase, which have been reported in EuZn$_2$Sb$_2$ even at temperatures greater than 100 K. This is further corroborated by finding weak temperature dependence of the MDC peak-splitting up to 130 K and over various incident photon energies. Interestingly, our ferromagnetic calculations, which reproduce our experimental bands, also generates a single pair of Weyl nodes above the Fermi level. Our study identifies EuZn$_2$Sb$_2$ as a minimal Weyl point candidate and an excellent platform for exploring the interplay between short-range magnetic interactions and electronic band structure.\\

\section*{Acknowledgements}
M.N. acknowledges support from the Air Force Office of Scientific Research MURI Grant No. FA9550-20-1-0322 and the 	National Science Foundation (NSF) CAREER Award No.
DMR-1847962. The work at Northeastern University was supported by the Air Force Office of Scientific Research under award number FA9550-20-1-0322 and benefited from the computational resource of Northeastern University's Advanced Scientific Computation Center (ASCC) and the Discovery Cluster. D.K. and T.R. were supported by the National Science Centre (Poland) under research grant 2021/41/B/ST3/01141. This work utilized resources of the SSRL at the SLAC National Accelerator Laboratory, supported by the U.S. Department 	of Energy, Office of Science, Office of Basic Energy Sciences under Contract No. DE-AC02-76SF00515. We thank Makoto Hashimoto and Donghui Lu for the beamline assistance at SSRL endstation 5-2. This research also used resources of the	ALS at the Lawrence Berkeley National Laboratory, a US Department of Energy Office of Science User Facility, under Contract No.	DE-AC02-05CH11231. We thank Jonathan Denlinger for beamline assistance at the ALS Beamline 4.0.3.

\clearpage

\setcounter{equation}{0}
\renewcommand{\theequation}{S\arabic{equation}}
\setcounter{figure}{0}
\renewcommand{\thefigure}{S\arabic{figure}}
\setcounter{section}{0}
\renewcommand{\thesection}{S\Roman{section}}
\setcounter{table}{0}
\renewcommand{\thetable}{S\arabic{table}}

\begin{center}
	\Large{\textbf{Supplemental Material: Observation of paramagnetic spin-degeneracy lifting in EuZn$_2$Sb$_2$}}\\[0.5 cm] \end{center}

\section{S1. Experimental and Computational Methods}
\subsection{Crystal growth and characterization}

Single crystals of EuZn$_2$Sb$_2$ were grown from Zn-Sb flux, following the procedure described elsewhere \cite{may2012properties}. High purity constituent elements (Eu 3N, Zn 4N, Sb 5N) were taken in the atomic ratio Eu:Zn:Sb of 1:5:5. The synthesis was carried out using an ACP-CCS-5 Canfield Crucible Set (LCP Industrial Ceramics Inc.) sealed in an evacuated quartz tube. The mixture was heated to 1000 $^{\circ}$C at a rate of 50 $^{\circ}$C/h, and maintained at this temperature for 20 hours. Subsequently, it was slowly cooled down to 800 $^{\circ}$C at a rate of 2 $^{\circ}$C/h, and then inverted and centrifuged to remove excess of molten ZnSb flux. As a product of this procedure, several black and shiny platelet-like crystals were obtained (FIG. S1(a)) with dimensions up to $3\times 2\times 0.7$ mm. The crystals were found to be stable against air and moisture.

Chemical composition and phase homogeneity of the prepared crystals were determined by energy-dispersive X-ray (EDX) analysis, performed using a FEI scanning electron microscope equipped with an EDAX Genesis XM4 spectrometer. The EDX results demonstrate single-phase character of the specimens and their targeted 1:2:2 stoichiometry.

In order to verify the crystal symmetry of the obtained single crystals, a small fragment was crumbled from a larger piece and examined on an Oxford Diffraction X'calibur four-circle single-crystal X-ray diffractometer equipped with a CCD Atlas detector. The experiment yielded a trigonal unit cell (space group $P\bar{3}m1$, No. 164) with lattice parameters $a$ = $b$ = 4.4873(5) $\text{\AA}$ and $c$ = 7.612(1) $\text{\AA}$, in good agreement with literature values \cite{may2012properties,weber2006low,zhang2008new}. Crystallinity and crystallographic orientation of the crystals used in physical measurements were determined by means of Laue X-ray backscattering implemented using a LAUE-COS (Proto) system (FIG. S1(b)). 
\subsection{Magnetic and electrical transport measurements}
EuZn$_2$Sb$_2$ crystals were further characterized for their physical behavior by means of magnetic and electrical transport measurements. Magnetic properties were investigated in the temperature range between 1.72 and 400 K and in magnetic fields up to 5 T using a Quantum Design MPMS-XL superconducting quantum interference device (SQUID) magnetometer. In this experiment, an external magnetic field was applied within the trigonal plane ($H||ab$). The electrical resistivity measurement was performed from 2 to 300 K using a Quantum Design PPMS-9 platform. The electrical leads were made of gold wiring attached to the bar-shaped specimens with silver epoxy paste. The experiments were done employing a standard four-point ac technique, and electrical current flowing within the crystallographic $ab$ plane ($i||ab$).

Magnetic measurements were performed on EuZn$_2$Sb$_2$ to assess the valency and ordering of the Eu magnetic moments. The results are shown in FIG. S1(c), where the sample is found to have antiferromagnetic (AFM) ordering at low temperatures. In the paramagnetic phase, the magnetic susceptibility follows a Curie-Weiss law with effective magnetic moment $\mu_{eff}$ = 7.88 $\mu_B$ and with a paramagnetic Curie temperature $\theta$ = 8.2 K. The value of $\mu_{eff}$ is close to the Russell-Saunders value of 7.94 $\mu_B$ for divalent Europium, while the positive sign of $\theta$ suggests predominant ferromagnetic exchange interaction. Nonetheless, as can be inferred from the upper inset to FIG S1(c), the compound undergoes an AFM phase transition at a Neel temperature of $T_N$ = 13.3 K, which is identified by the presence of a cusp-like maximum in $\chi (T)$. Another indication of the AFM ordering in EuZn$_2$Sb$_2$ comes from the behavior of the magnetization as a function of applied magnetic field at temperatures much below $T_N$. The lower inset to FIG S1(c) shows the magnetization isother measured at $T$ = 1.72 K, which exhibits a linear field dependence in magnetic fields smaller than $\mu_0 H^*$ = 3.4 T, and a plateau in stronger fields. The saturated magnetization is 70 emu/g, which corresponds to 6.59 $\mu_B$ magnetic moments, somewhat smaller than the expected 7 $\mu_B$ for divalent Eu$^{2+}$ ions. The $\sigma (H)$ variation is fully reversible, demonstrating the lack of hysteresis.
\subsection{Spectroscopic measurements}
High-resolution angle-resolved photoemission spectroscopy (ARPES) measurements on EuZn$_2$Sb$_2$ were performed at the Stanford Synchrotron Radiation Laboratory (SSRL) endstation 5-2 and at the Advanced Light Source (ALS) beamline 4.0.3, equipped with Scienta DA30 analyzer and Scienta R8000 electron analyzer, respectively. The pressure was maintained to better than 5$\times$10$^{-11}$ Torr during measurements. EuZn$_2$Sb$_2$ single crystals were cleaved in-situ, exposing the (001) surface used for ARPES measurement. The energy resolution was set to 20 meV at the SSRL and 30 meV at ALS. Measurements at the SSRL were performed at a temperature of 20 K.
 
\subsection{Computational details}

All electronic structure calculations were performed within the density functional theory (DFT) framework using a plane wave basis set in the Vienna Ab initio Simulation Package (VASP) \cite{kresse1996efficient}. The projector-augmented-wave (PAW) pseudopotentials were used \cite{kresse1999ultrasoft}. The kinetic energy cutoff for the plane wave basis was set to 400 eV. The Brillouin zone integration was performed using a 8$\times$8$\times$4 k-mesh. The screened hybrid functional developed by Heyd-Scuseria-Ernzerhof (HSE) was used for obtaining the band structure \cite{heyd2003hybrid}. The screening parameter was set to 0.2 (HSE06) \cite{heyd2005energy}. The lattice parameters used in the calculations are $a$ = $b$ = 4.487 $\text{\AA}$ and $c$ = 7.612 $\text{\AA}$.

\section{S2. Band structure along $\mathrm{\overline{M}}-\mathrm{\overline{\Gamma}}-\mathrm{\overline{M}}$ at different photon energies}

Band-dispersion perpendicular to the cleaved surface was obtained through varying the incident photon energy at SSRL endstation 5-2. Several $\mathrm{\overline{M}}-\mathrm{\overline{\Gamma}}-\mathrm{\overline{M}}$ cuts were taken after increasing the incident photon energy by 2 eV increments between 20 eV and 120 eV. The results are presented in FIG. S2. The overall band structure looks similar at the measured photon energies. Importantly, the presence of the split band is seen in all of the cuts, establishing that the ferromagnetic fluctuation induced splitting exists in this material.

\begin{figure*}
	\centering
	\includegraphics[width=1\linewidth]{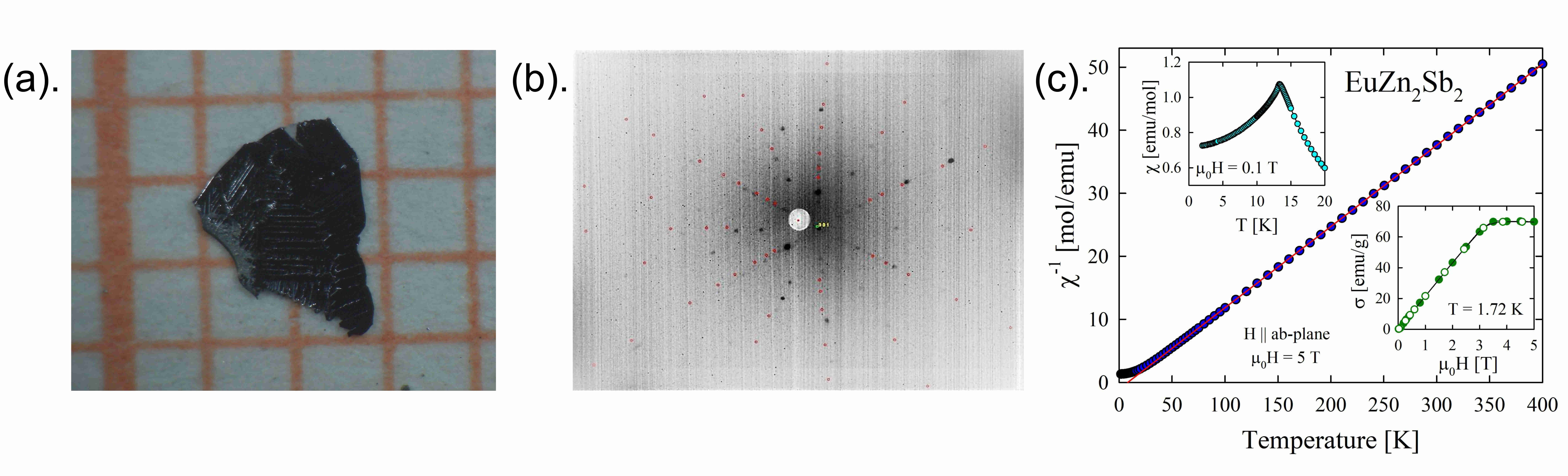}
	\caption{Crystallographic and magnetic characterization of EuZn$_2$Sb$_2$ single crystals. (a) Picture of the EuZn$_2$Sb$_2$ sample. (b) Laue X-ray backscattering diffraction pattern. (c) Inverse magnetic susceptibility over temperatures of 1.72 to 400 K with a 5 T in-plane applied field (main). Magnetic susceptibility at low temperatures for an applied field of 0.1 T (top inset). Magnetization over the applied field taken at a constant temperature of 1.72 K (lower inset).}
	\label{fig:eu12sb2-supfig-1-04-25-2023}
\end{figure*}

\begin{figure*}
	\centering
	\includegraphics[width=0.9\linewidth]{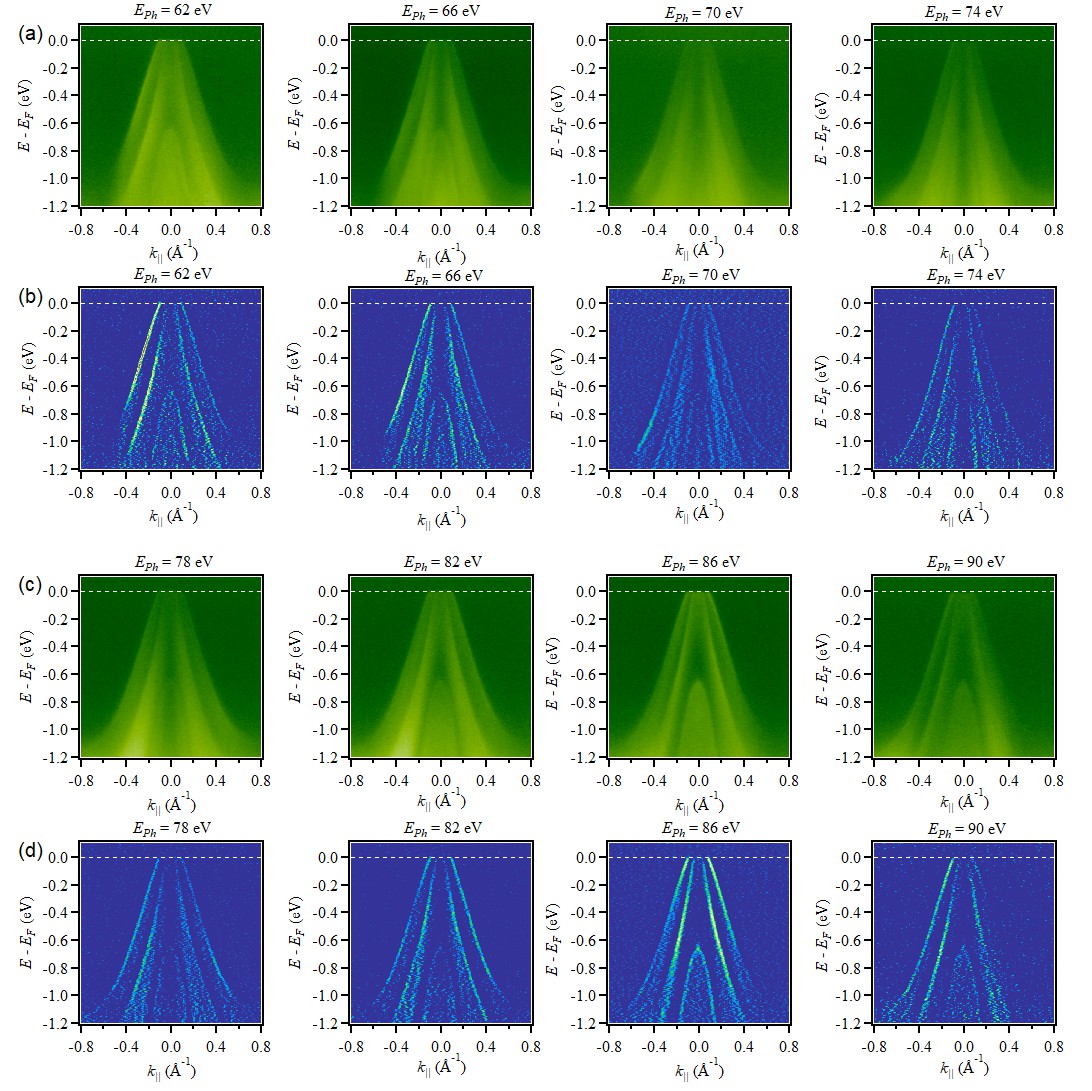}
	\caption{Incident photon energy variation. (a) $\mathrm{\overline{M}}-\mathrm{\overline{\Gamma}}-\mathrm{\overline{M}}$ cuts taken for incident photon energies between 62 and 74 eV. (b) Second derivative plots for corresponding figures in (a). (c) Cuts taken for incident photon energies between 78 and 90 eV. (d) Second derivative plots for corresponding figures in (c).}
	\label{fig:eu122-supfig-2-04-27-23}
\end{figure*}


\section{S3. MDC Fitting Procedure}

Curve fits of the momentum distribution curves (MDCs) were performed using Voigt functions. ARPES provides a direct measurement of the single-particle spectral function, which for a weakly interacting system takes the form of a Lorentzian at a constant energy in momentum space \cite{damascelli2003angle}. Temperature and resolution limitations convolute this theoretical Lorentzian lineshape with a Gaussian function, justifying the use of the Voigt function for fitting of the MDCs \cite{levy2014deconstruction}. First a rough fit is performed using Gaussian functions, where the function baseline, noise reduction, and a minimum peak fraction are specified in order to reject false peaks. Once we no longer get false peaks from the noise baseline, we repeat the fit with Voigt functions to characterize the MDC lineshapes originating from the bands. We show the results of this multi-peak fit in FIG. 3.(b,c) of the main text at binding energies of 0.5 eV and 0.4 eV along the  $\mathrm{\overline{M}}-\mathrm{\overline{\Gamma}}-\mathrm{\overline{M}}$ direction, respectively. 

\section{S4. $\mathrm{\overline{M}}-\mathrm{\overline{\Gamma}}-\mathrm{\overline{M}}$ band structure with varying Sample Temperature}

Temperature dependence was obtained at the ALS beamline 4.0.3. For this we oriented the analyzer slit along the $\mathrm{\overline{M}}-\mathrm{\overline{\Gamma}}-\mathrm{\overline{M}}$ high symmetry direction and took repeated cuts at varying temperatures. The temperature was increased in increments of 20 K between 30 K and 130 K. Temperature variations of the MDCs taken at a binding energy of 0.5 eV are stacked in FIG 3.(e) of the main text. We find weak temperature dependence of the MDC lineshapes across these temperatures. Within the expected variations in the fitting we find no change in the splitting size nor in the relative intensities of the split bands upon accounting for gaussian-temperature broadening. 

\section{S5. Electronic structure studied under different photon polarizations}

ARPES maps were taken for both linear vertical and linear horizontal polarizations of 86 eV incident photons at SSRL endstation 5-2. To check for matrix-element suppression of the valence band structure, we utilize both linear vertical (p) and horizontal (s) polarization of the incident photons, which selects for Bloch-states that possess odd and even reflection symmetry about the photoemission plane, respectively. Strong suppression of the photoemission intensity is observed from the valence band structure upon using horizontal-polarized incident photons near the $\Gamma$ point (FIG. S3). 
\begin{figure*}
	\centering
	\includegraphics[width=0.7\linewidth]{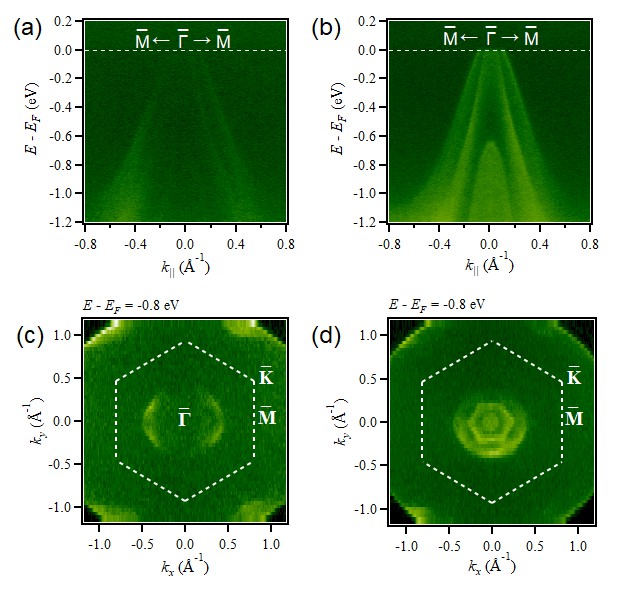}
	\caption{Comparison of linear-vertical (LV) and linear-horizontal (LH) incident light polarization in EuZn$_2$Sb$_2$. MGM dispersion (a) LV and (b) LH, corresponding to p and s-polarization in the experimental geometry, respectively. 0.8 eV constant energy cut for (c) LV and (d) LH at a binding energy of 0.8 eV.}
	\label{fig:eu12sb2-figure-4-01-26-2023}
\end{figure*}
\section{S6. Band Structure calculations using GGA and HSE functionals}
\begin{figure*}
	\centering
	\includegraphics[width=0.7\linewidth]{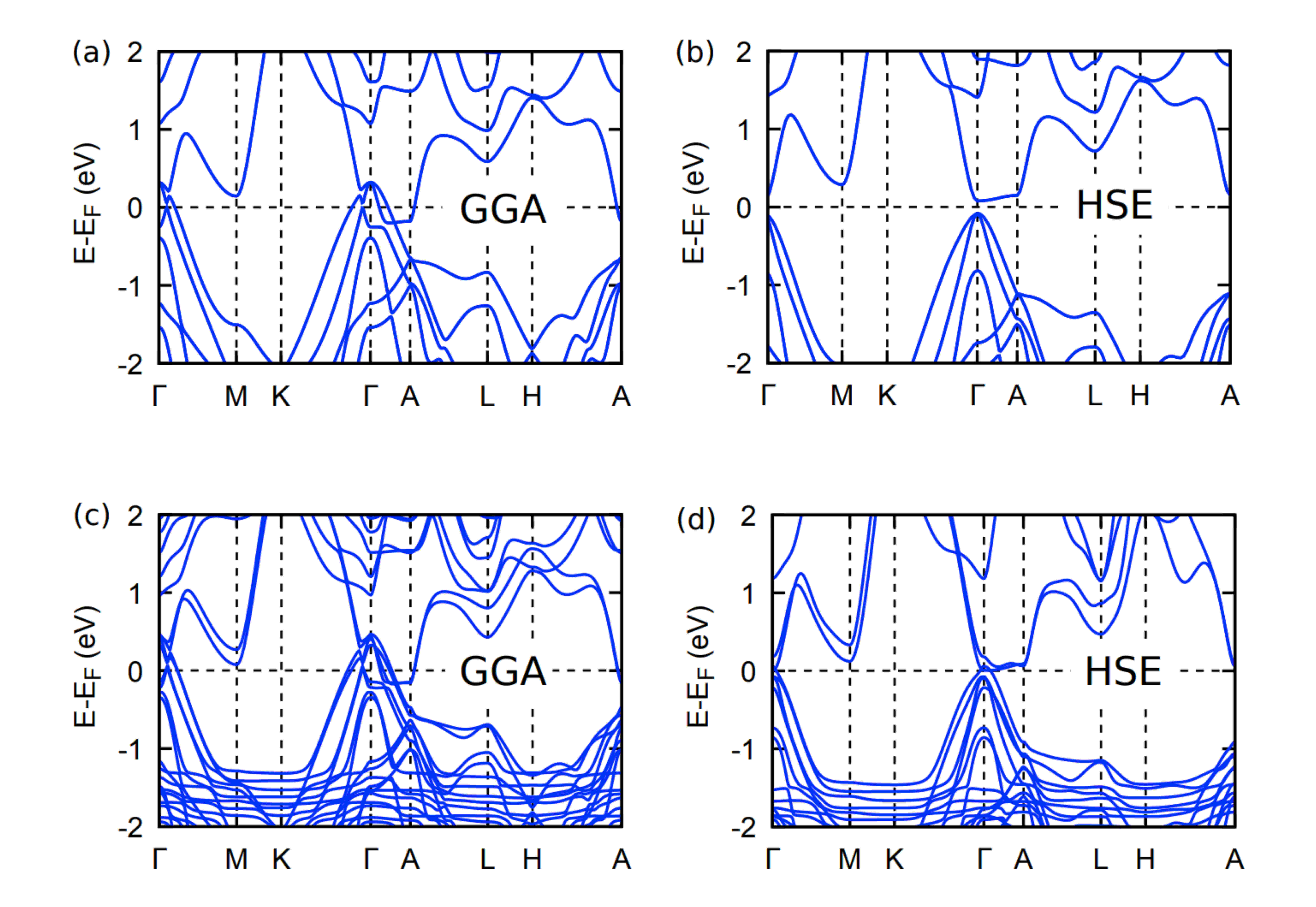}
	\caption{Calculated band dispersion using GGA and HSE functionals. (a,b) Comparison of the GGA and HSE computed band structure for the NM configuration. (c,d) Comparison of the GGA and HSE computed band structure for the out-of-plane ferromagnetic configuration. The electronic structure computed using the HSE functional better fits our experimental data. Note that, unlike in the main text, the Fermi levels are not shifted in these panels.}
	\label{fig:bgsupplementary}
\end{figure*}
First-principles calculations were performed using both the generalized gradient approximation (GGA) and the hybrid HSE functionals. GGA-based calculations failed to replicate the experimental band structure of EuZn$_2$Sb$_2$, predicting highly metallic electronic structure at the Fermi level (FIG. S4(a)). Band dispersion calculations were replicated using the HSE functional, which resulted in a simplified valence band structure consistent with ARPES measurements (FIG. S4(b)). Out-of-plane ferromagnetic calculations were performed using both functionals where the spin-polarization of the valence states are observed in both results. However, the HSE-based calculation predicts a trivial insulator-to-Weyl semimetal transition upon magnetization of the sample.

\section{S7. Projected weyl nodes in FM calculations}


Ferromagnetic calculations were performed with an out-of-plane orientation of the Eu moments. The result of these calculations not only replicate the valence band structure observed experimentally, but also project two Weyl points located just below the theoretical Fermi level. Due to the experimentally determined Fermi level being shifted down by about 180 meV from the theoretical Fermi level, these Weyl nodes are expected to reside well above the Fermi level in our sample. To confirm the crossings to be topologically non-trivial, we calculated the Wannier charge center evolution around a sphere enclosing the gapless points (FIG 3(h) of the main text). This calculation assigned a chirality of $\pm{1}$ to the Weyl nodes.


\end{document}